\documentstyle[preprint,eqsecnum,aps,epsf]{revtex}
\newcommand{\gsim}{\mbox{\raisebox{-1.ex}{$\stackrel
     {\textstyle>}{\textstyle\sim}$}}}
\newcommand{\lsim}{\mbox{\raisebox{-1.ex}{$\stackrel
     {\textstyle<}{\textstyle \sim}$}}}

\newcommand{\beq}{\begin{equation}}
\newcommand{\eeq}{\end{equation}}
\newcommand{\beqa}{\begin{eqnarray}}
\newcommand{\eeqa}{\end{eqnarray}}
\newcommand{\bea}{\begin{array}}
\newcommand{\ena}{\end{array}}
\begin{document}
\draft
\title{Particle velocity in noncommutative space-time}
\author{Takashi Tamaki
\thanks{electronic mail:tamaki@gravity.phys.waseda.ac.jp},
Tomohiro Harada,\thanks{electronic mail:harada@gravity.phys.waseda.ac.jp}
Umpei Miyamoto,\thanks{electronic mail:umpei@gravity.phys.waseda.ac.jp}
}
\address{Department of Physics, Waseda University,
Ohkubo, Shinjuku, Tokyo 169-8555, Japan}
\author{and \\ 
Takashi Torii
\thanks{electronic mail:torii@gravity.phys.waseda.ac.jp}}
\address{
Advanced Research Institute for Science and Engineering,
Waseda University, Ohkubo,
Shinjuku, Tokyo 169-8555, Japan}
\date{\today}
\maketitle
%------------------------------
\begin{abstract}
We investigate a particle velocity in the $\kappa$-Minkowski space-time, 
which is one of the realization of a noncommutative space-time. 
We emphasize that arrival time analyses by high-energy $\gamma$-rays or 
neutrinos, which have been considered as powerful tools to restrict the violation 
of Lorentz invariance, are not effective to detect space-time noncommutativity. 
In contrast with these examples, we point out a possibility that 
{\it low-energy massive particles} play 
an important role to detect it.
\end{abstract}

%%%%%%%%%%%%%%%%%%%%%%%%%%%%%%%%%%%%%%%
%%%%%%%%%%%%%%%%%%%%%%%%%%%%%%%%%%%%%%%
\section{Introduction}
%%%%%%%%%%%%%%%%%%%%%%%%%%%%%%%%%%%%%%%
%%%%%%%%%%%%%%%%%%%%%%%%%%%%%%%%%%%%%%%

It is believed that general relativity describes the large scale structure
of space-time, and it has revealed the history 
and present states of our universe. The direct evidence for the validity of general
relativity in strong gravitational regime will be obtained by the
observations of gravitational waves from inspiraling
binaries in near future. While little is known about the small scale
structure of space-time because gravity should be also quantized consistently, 
which is not completed yet, in such a regime. The physics of small scale structure is important
because there are some phenomena in our universe which such physics may 
be needed to describe, e.g., the birth of the universe, space-time
singularity and ultra-high energy cosmic rays. The last one is one of the main 
topics of this paper.

Although we do not have the coherent theory of  small scale structure, 
several attempts have been made
to extract its information and effects. Among them, the simplest
way is modifying the dispersion relation that leads to the
violation of Lorentz invariance. We call the theories obtained by
this method modified dispersion relation (MDR) models.

The violation of Lorentz invariance appears in the context
of string/M theories, where the space-time structure is modified to include
the space-time noncommutativity\cite{Connes}. The space-time noncommutativity
also arises as a result of deformation quantization\cite{Jimbo}.
By using the MDR models, the robustness of the spectrum of a black hole
evaporation and of the fluctuation generated in an inflationary cosmology were
discussed\cite{Unruh,Branden}. Remarkably enough, it was discussed that 
the anomalous detection of the ultra-high energy cosmic rays 
beyond the Greisen-Zatsepin-Kuzmin 
cutoff may be observational evidence for the violation of Lorentz
invariance\cite{LIVGZK,Tamaki}.

It has been discussed that there is a severe constraint on the energy scale where
the Lorentz invariance might be violated, i.e., the scale of quantum
gravity as $E_{QG}\gsim 7.2\times 10^{16}$ GeV by the detection of 
$\gamma$-rays from Markarian (Mk) 421 as pointed out in
Refs.~\cite{Biller,Nature,Garay,Camelia}. They examined the energy
dependence of the arrival time of photons
and compared it with the observations of Mk 421 to obtain this constraint.

In general, however, it is plausible that not only a dispersion relation but
also other relations such as energy-momentum conservation laws
might be altered in a Planck scale physics.
A particle velocity in such models may have qualitative difference from 
that in the MDR models. To investigate these features, we employ a model called
$\kappa$-Minkowski space-time, where noncommutativity is introduced as
$[x^{i},t]=i\lambda x^{i}$ \cite{MandR,Zakr,LRZ}, and 
compare a group velocity in the $\kappa$-Minkowski space-time
evaluated in our previous paper\cite{Tamaki} with that in the MDR models. 
The properties of this group velocity were also investigated in Ref.~\cite{LandN}. 

This paper is organized as follows. In Sec. II, we review the previous 
discussion in the MDR models. After introducing the $\kappa$-Minkowski space-time 
in Sec. III, we discuss the particle velocity in this model in Sec. IV. 
In Sec. V, we consider the observational possibilities of time delay by comparing 
a particle velocity in the usual Minkowski space-time with that in 
the $\kappa$-Minkowski space-time and MDR models. 
We show that the space-time noncommutativity
does {\it not} affect the velocity of massless particles,
which implies that the arrival time analysis by $\gamma$-rays
is {\it not} useful to detect the space-time noncommutativity. 
We also discuss a possibility that
the space-time noncommutativity might be detected by using low-energy massive
particles. In Sec. VI, we summarize our results and mention future work. 
We use the signature $(-,+,+,+)$ and units in which $c=\hbar =1$ below.

%%%%%%%%%%%%%%%%%%%%%%%%%%%%%%%%%%%%%%%
%%%%%%%%%%%%%%%%%%%%%%%%%%%%%%%%%%%%%%%
\section{Modified Dispersion Relation Models}
%%%%%%%%%%%%%%%%%%%%%%%%%%%%%%%%%%%%%%%
%%%%%%%%%%%%%%%%%%%%%%%%%%%%%%%%%%%%%%%

Although there are various ways to modify the dispersion
relation, we consider here the form in Ref. \cite{Nature}
as $p^{2}+m^{2}=E^{2}[1+f(E/E_{QG})]$, where
$f$ is a model-dependent function and $E_{QG}$ is the
effective energy scale of quantum gravity. For simplicity, 
we assume that $f$ is an analytic function. 
Although in general, $f$ and $E_{QG}$ may depend on the
species and properties of the particles\cite{LIVGZK},
we do not consider this possibility, which implies that the
effects of  quantum gravity originate from the space-time structure.
In the low-energy limit, $E\ll E_{QG}$, the above dispersion
relation becomes
%%%%%%%%%%%%%%%
\begin{eqnarray}
p^{2}+m^{2}=E^{2}+\frac{\xi E^{n}}{E_{QG}^{n-2}},
\label{LIV}
\end{eqnarray}
%%%%%%%%%%%%%%%
up to the lowest correction. We have chosen $\xi =\pm 1$ and $n\geq 3$ 
is the integer, 
which is determined by the form of the function $f$.
Note that $E<m$ for $\xi =1$ in the low-momentum limit.
This type of dispersion relation also appears in the
Liouville string approach to
quantum gravity\cite{Ellis}.

The velocity $v_{MDR}$ in this model is obtained by
differentiating the dispersion relation (\ref{LIV}) with
respect to $p$, 
%%%%%%%%%%%%%%%
\begin{eqnarray}
v_{MDR}:=\frac{dE}{dp}= \frac{2\sqrt{E^{2}-m^{2}+
\xi E^{n}/E_{QG}^{n-2}}}{2E+n\xi E^{n-1}/E_{QG}^{n-2}}.
\label{vLIV}
\end{eqnarray}
%%%%%%%%%%%%%%%
It should be noted that $v_{MDR}$ depends on the energy even for 
massless particles because of the correction term.
We can make use of the energy dependence to restrict $E_{QG}$.

Let us consider a $\gamma$-ray from the distant source.
We approximate the velocity of the $\gamma$-ray 
by expanding Eq.~(\ref{vLIV}) by $E/E_{QG}$ to
%%%%%%%%%%%%%%%
\begin{eqnarray}
v_{MDR} &\approx & 1
-\frac{\xi (n-1)}{2}\left(\frac{E}{E_{QG}}\right)^{n-2}.
\label{first2adv}
\end{eqnarray}
%%%%%%%%%%%%%%%
Although the correction term may be very small, the difference
of arrival time depending on the energy of the photons may become large
enough to measure if the $\gamma$-rays travel a very long distance
\cite{Biller,Nature,Garay,Camelia}.
The time delay is evaluated as
%%%%%%%%%%%%%%%
\begin{eqnarray}
\delta t &=& \frac{L}{v_{MDR}(E_1)}-\frac{L}{v_{MDR}(E_2)}
\nonumber \\
&\approx & \frac{(n-1)\xi L}{2{E_{QG}}^{n-2}}
({E_{1}}^{n-2}-{E_{2}}^{n-2}),
\end{eqnarray}
%%%%%%%%%%%%%%%
where $L$, $E_{1}$ and $E_{2}$ are the distance from the source to the
Earth, amounts of the energy of particles $1$ and $2$, respectively.

One of the examples of this kind of analyses is the arrival time analysis 
by $\gamma$-rays from Mk 421 ($\sim 150$ Mpc from the Earth).
It was reported that $\gamma$-rays in the energy range between
1 and 2 TeV arrived at the Earth within the time difference 
$\sim 200$ seconds\cite{Biller}.
Then, $E_{QG}$ is  constrained to $E_{QG}$
$\agt \left[ 3.6\times (n-1)(n-2)\times 10^{13}\right]^{1/(n-2)}\times 10^{3}$ GeV.
Since the value of $n$ has been assumed to be 3 in most of the previous
works, it has been concluded that $E_{QG}\agt 7.2 \times 10^{16}$ GeV.
We should note, however,
that $n$ may be 4 or larger. In this case, the constraint becomes
$E_{QG}\agt 1.5 \times 10^{10}$ GeV for $n=4$ and 
$E_{QG}\agt 7.6 \times 10^{7}$ GeV for $n=5$.
Hence the constraint may become quite loose compared with 
the previous reports.

%%%%%%%%%%%%%%%%%%%%%%%%%%%%%%%%%%%%%%%
%%%%%%%%%%%%%%%%%%%%%%%%%%%%%%%%%%%%%%%
\section{$\kappa$-Poincar\`{e} algebra and
$\kappa$-Minkowski space-time}
%%%%%%%%%%%%%%%%%%%%%%%%%%%%%%%%%%%%%%%
%%%%%%%%%%%%%%%%%%%%%%%%%%%%%%%%%%%%%%%

Here, we review the $\kappa$-Poincar\`{e}
algebra\cite{Lukierski}, which has the structure of a Hopf
algebra (quantum group)\cite{majid}. The generators of the
$\kappa$-Poincar\`{e} algebra $\cal{P}_{\kappa}$ satisfy the following
commutation relations:
%%%%%%%%%%%%%%%
\begin{eqnarray}
\left[M_{\mu\nu}, M_{\rho\sigma}\right] &=& i\left(\eta
_{\mu\sigma}M_{\nu\rho}-\eta _{\mu\rho}M_{\nu\sigma}+\eta
_{\nu\rho}M_{\mu\sigma}-\eta _{\nu\sigma}M_{\mu\rho}\right), \\
\left[M_{i}, p_{0}\right] &=& 0, \\
\left[M_{i}, p_{j}\right] &=& i\epsilon_{ijk}p_{k}, \\
\left[N_{i}, p_{0}\right] &=& ip_{i}, \\
\left[N_{i}, p_{j}\right] &=& -i\delta_{ij}\left[\frac{1}{2\lambda}
\left(1-e^{2p_{0}\lambda}\right)
+\frac{\lambda}{2}\mbox{\boldmath $p$}^{2}\right]+i\lambda p_{i}p_{j},\\
\left[p_{\mu}, p_{\nu}\right] &=& 0,
\end{eqnarray}
%%%%%%%%%%%%%%%
where $M_{i}\equiv\frac{1}{2}\epsilon _{ijk}M_{jk},N_{i}\equiv M_{0i}$ and
$p_{\mu}$ are generators of rotation, boost and 
translation, respectively. The Greek and Roman indicies take the values from $0$ to $3$ 
and from $1$ to $3$, respectively. We abbreviate $\sum_{i}p_{i}^{2}$ as 
$\mbox{\boldmath $p$}^{2}$. We can recover the ordinary 
commutation relations of the Poincar\`{e} algebra in the limit $\lambda\to 0$. 
The dispersion relation is determined by the eigenvalue of the Casimir operator 
that commutes with all elements in $\cal{P}_{\kappa}$:
%%%%%%%%%%%%%
\begin{eqnarray}
\frac{2\cosh(\lambda p_{0})}{\lambda^{2}}
-\mbox{\boldmath $p$}^{2}e^{-\lambda
p_{0}}=\frac{2\cosh(\lambda m)}{\lambda^{2}},
\label{dis}
\end{eqnarray}
%%%%%%%%%%%%%
where the rest mass $m$ is defined as the energy with $p_{i}=0$. 
  The coproducts $\Delta: \cal{P}_{\kappa}\rightarrow
\cal{P}_{\kappa}\otimes
\cal{P}_{\kappa}$ of the basic generators are
%%%%%%%%%%%%%%%
\begin{eqnarray}
\Delta(M_{i})&=&M_{i}\otimes 1+1\otimes M_{i},
\\
\Delta(N_{i})&=&N_{i}\otimes 1+e^{p_{0}\lambda}\otimes N_{i}
-\lambda\epsilon
_{ijk} p_{j}\otimes M_{k},\\
\Delta(p_{0})&=&p_{0}\otimes 1+1\otimes p_{0}, \label{eq:energy}
\\
\Delta(p_{i})&=&p_{i}\otimes 1+e^{p_{0}\lambda}\otimes
p_{i}. 
\label{eq:momentum}
\end{eqnarray}
%%%%%%%%%%%%%%%
The above coproducts of $p_{\mu}$, (\ref{eq:energy}) and
(\ref{eq:momentum}), are interpreted as the non-Abelian addition law of
energy-momenta for particles 1 and 2 as
%%%%%%%%%%%%%%%
\begin{eqnarray}
 \left(E_{1},\mbox{\boldmath $p$}_{1}\right)
 \hat{+}\left(E_{2},\mbox{\boldmath $p$}_{2}\right)
 :=\left(E_{1}+E_{2},\mbox{\boldmath $p$}_{1}
 +e^{\lambda E_{1}} \mbox{\boldmath $p$}_{2}\right),
 \label{add}
\end{eqnarray}
%%%%%%%%%%%%%%%
where we identify $p_{0}$ with energy $E$. Note that the associativity of 
the addition law is given by the coassociativity $(\Delta\otimes
id)\circ\Delta=(id\otimes\Delta)\circ\Delta$. The coproducts of other
elements in $\cal{P}_{\kappa}$ are extended as $\Delta(1)=1\otimes 1$ and
$\Delta(MM')=\Delta(M)\Delta(M'), \forall M,M' \in \cal{P}_{\kappa}$. We can
check the consistency between this extension of the coproducts as an algebra
homomorphism and the commutation relation, i.e.,  $\Delta\left[M,
M'\right]=\left[\Delta M, \Delta M'\right]$. This consistency guarantees
that the $\kappa$-Poincar\`{e} algebra is form-invariant for multi-particle
systems.

The asymmetry of the coproducts for the permutation of particles is called
noncocommutativity.
The noncocommutativity of the coproducts for the translation sector $T
\subset \cal{P}_{\kappa}$ has two important meanings.
One is that the noncommutativity of the $\kappa$-Minkowski space-time
is a direct consequence of the noncocommutativity. Elements in the $\kappa$-Minkowski
space-time are defined as linear functionals on the translation
sector, $T^{\ast}:T\rightarrow \mathbf{C}$. The products in $T^{\ast}$ is
defined in terms of coproducts in $T$, i.e., $\forall x, y \in T^{\ast}$ and
$\forall p \in T$,
%%%%%%%%%%%%%%%
 \begin{eqnarray}
\langle xy, p \rangle :=&&\langle x\otimes y, \Delta p \rangle \\
=&&\sum _{a}\langle x, p_{a (1)}\rangle \langle y,p_{a (2)}\rangle,
\end{eqnarray}
%%%%%%%%%%%%%%%
where we write the coproducts as $\Delta(p)=\sum _{a}p_{a (1)}\otimes p_{a
(2)}$.
With the duality relations $\langle x^{\mu},
p_{\nu}\rangle=-i\delta^{\mu}_{\nu}$, this leads to the following
commutation relations\cite{MandR}:
%%%%%%%%%%%%%%%
\begin{eqnarray}
\left[x^{i}, x^{0}\right] & =&i\lambda x^{i}, \\
\left[x^{0}, x^{0}\right]&=&0, \\
\left[x^{i}, x^{j}\right]&=&0.
\end{eqnarray}
%%%%%%%%%%%%%%%
The other is that the noncocommutativity leads to a deformed
group velocity formula\cite{Tamaki}, which is different from
the usual velocity formula
$dE/d\mbox{\boldmath $p$}$ as will be shown in the next section.

We can also define differentiation, integration and Fourier 
transformation\cite{Oeckl}. 
The plane wave $\psi _{(E,\mbox{\boldmath $p$})}=e^{i\mbox{\boldmath
$p$}\cdot \mbox{\boldmath $x$} }e^{iEt}$ in the $\kappa$-Minkowski space-time
introduced in \cite{1plane,CandM} respects the non-Abelian addition law of
energy-momenta in the sense 
%%%%%%%%%%%%%
\begin{eqnarray}
\psi _{(E_{1},\mbox{{\small\boldmath $p$}}_{1})}\psi _{(E_{2}, {\small
\mbox{\boldmath$p$}}_{2})} &=& e^{i{\small\mbox{\boldmath $p$}}_{1}
\cdot {\small\mbox{\boldmath$x$}}}e^{iE_{1}t}e^{i{\small\mbox{\boldmath $p$}}_{2}
\cdot {\small\mbox{\boldmath $x$}} }e^{iE_{2}t} \\
 &=&\psi _{(E_{1}+E_{2}, {\small\mbox{\boldmath $p$}}_{1}+e^{\lambda
E_{1}}{\small\mbox{\boldmath $p$}}_{2})}.
\end{eqnarray}
%%%%%%%%%%%%%

%%%%%%%%%%%%%%%%%%%%%%%%%%%%%%%%%%%%%%%
%%%%%%%%%%%%%%%%%%%%%%%%%%%%%%%%%%%%%%%
\section{Velocity formula}
%%%%%%%%%%%%%%%%%%%%%%%%%%%%%%%%%%%%%%%
%%%%%%%%%%%%%%%%%%%%%%%%%%%%%%%%%%%%%%%

From the properties in the $\kappa$-Minkowski space-time in Sec~III,
we can establish group velocity formulae.
For this purpose, we consider infinitesimal changes $\Delta E$
and $\Delta \mbox{\boldmath $p$}$ in $E$ and $\mbox{\boldmath $p$}$,
respectively, as a result of adding
($\Delta E',\Delta \mbox{\boldmath $p$}'$) as
%%%%%%%%%%%%%%%
\begin{eqnarray}
(E,\mbox{\boldmath $p$})\hat{+}(\Delta E',\Delta\mbox{\boldmath $p$}')
=(E+\Delta E,\mbox{\boldmath $p$}+\Delta \mbox{\boldmath $p$}).
\label{eq:9}
\end{eqnarray}
%%%%%%%%%%%%%%%
By the addition law (\ref{add}), we have 
%%%%%%%%%%%%%%%
\begin{eqnarray}
(\Delta E',\Delta \mbox{\boldmath $p$}')=(\Delta E,\frac{\Delta
\mbox{\boldmath $p$}}{e^{\lambda E}}) .
\label{eq:9d}
\end{eqnarray}
%%%%%%%%%%%%%%%
Next, we construct a wave packet by superposing plane
waves. Here we only consider two waves for simplicity, whose momenta and 
amounts of energy are different infinitesimally from each other\cite{footnote2}.
%%%%%%%%%%%%%%%
\begin{eqnarray}
I&=&\psi_{(E-\Delta E, {\small\mbox{\boldmath $p$}}-\Delta
{\small\mbox{\boldmath $p$}})}
+\psi_{(E+\Delta E, {\small\mbox{\boldmath $p$}}
+\Delta {\small\mbox{\boldmath $p$}})}
\nonumber \\
&\cong &2e^{i{\small\mbox{\boldmath $p$}}
\cdot{\small\mbox{\boldmath $x$}}}
e^{iEt}\cos \left[\frac{\Delta \mbox{\boldmath $p$}}{e^{\lambda E}}
\cdot
\left(\mbox{\boldmath $x$}+\frac{e^{\lambda E}\Delta Et}{\Delta
\mbox{\boldmath $p$}}\right)\right] ,\label{eq:11}
\end{eqnarray}
%%%%%%%%%%%%%%% 
where we neglected the terms that vanish in the limit $\Delta
\mbox{\boldmath$p$}\to 0$.
The group velocity $\mbox{\boldmath $v$}_{l}$ of this wave packet
can be written as 
%%%%%%%%%%%%%%%
\begin{eqnarray}
\mbox{\boldmath $v$}_{l}:=e^{\lambda E}\frac{dE}{d\mbox{\boldmath $p$}}.
\label{eq:12}
\end{eqnarray}
%%%%%%%%%%%%%%%
There remains ambiguity in constructing the wave packet because of the
noncommutativity of the space-time. Another possibility is
%%%%%%%%%%%%%%%
\begin{eqnarray}
(\Delta E',\Delta\mbox{\boldmath $p$}')\hat{+}(E,\mbox{\boldmath $p$})
=(E+\Delta E,\mbox{\boldmath $p$}+\Delta \mbox{\boldmath $p$}). 
\label{eq:9c}
\end{eqnarray}
%%%%%%%%%%%%%%%
In this case, the corresponding group velocity
$\mbox{\boldmath $v$}_{r}$ is
%%%%%%%%%%%%%%%
\begin{eqnarray}
\mbox{\boldmath $v$}_{r}:=\left(
1-\lambda {\small\mbox{\boldmath $p$}}\cdot\frac{dE}{
d{\small\mbox{\boldmath $p$}}} \right)^{-1}
\frac{dE}{d{\small\mbox{\boldmath $p$}}}.
\label{eq:12c}
\end{eqnarray}
%%%%%%%%%%%%%%%
These velocities can be expressed explicitly in terms of the functions
of $E$ and $m$ by using the dispersion relation.
By the definitions of $\mbox{\boldmath $v$}_{l}$ and
$\mbox{\boldmath $v$}_{r}$, we find
%%%%%%%%%%%%%%%
\begin{eqnarray}
\mbox{\boldmath $v$}_{l}&=&\frac{e^{\lambda E/2}
\sqrt{2[\cosh (\lambda E)-\cosh (\lambda m)]}}{|e^{\lambda E}-
\cosh (\lambda m)|}\mbox{\boldmath $e$},\label{vl-vector}\\
\mbox{\boldmath $v$}_{r}&=&\frac{e^{-\lambda E/2}
\sqrt{2[\cosh (\lambda E)-\cosh (\lambda m)]}}{|e^{-\lambda E}-
\cosh (\lambda m)|}\mbox{\boldmath $e$},
\label{vr-vector}
\end{eqnarray}
%%%%%%%%%%%%%%%
where $\mbox{\boldmath $e$}:=
\mbox{\boldmath $p$}/|\mbox{\boldmath $p$}|$. We find that 
the velocities have the same direction as that of the momenta.
Note also that there is a 
correspondence between the transformations $\lambda \to -\lambda$ and
$\mbox{\boldmath $v$}_{l} \to \mbox{\boldmath $v$}_{r}$.

These velocities were also investigated by Lukierski and Nowicki and the following  
facts were pointed out in Ref. \cite{LandN}: (i) $v_{l}:=|\mbox{\boldmath $v$}_{l}|$, 
$v_{r}:=|\mbox{\boldmath $v$}_{r}|\leq 1$ 
for all energies, (ii) $dv_{l}/dE>0$, $dv_{r}/dE>0$, and 
(iii) $v_{r}$ has a classical velocity addition law, i.e., the addition of parallel 
velocities $v_{r1}$ and $v_{r2}$ becomes 
%%%%%%%%%%%%%%%
\begin{eqnarray}
v_{r12}=\frac{v_{r1}+v_{r2}}{1+v_{r1}v_{r2}} .
\label{v-addition}
\end{eqnarray}
%%%%%%%%%%%%%%%
If the boost generator $N_{i}$ was an even function for $\lambda$, 
this addition law would hold even for $v_{l}$ because of the correspondence mentioned 
above. However, this is not the case. We postpone the interpretation of 
this asymmetry as future work. 

%%%%%%%%  application
Next, we discuss the application of the above velocity formulae. 
In the MDR models, since the energy scale of 
quantum gravity  $E_{QG}$ is introduced
perturbatively (see Eq~(\ref{first2adv})),
it is reasonable to apply the velocity formulae under
the condition on $E\ll E_{QG}$. While if we apply the velocity 
formulae in the $\kappa$-Minkowski space-time, the energy range 
is not restricted.

Let us examine the case beyond the quantum gravity 
scale, i.e., $|\lambda E|\gg 1$. Since we can obtain the information about 
$v_{r}$ by using the transformation $\lambda\to -\lambda$ to $v_{l}$, 
we only examine $v_{l}$ below. We evaluate the velocity $v_{l}$ in the 
following limits (see Table \ref{table}.). 
When $\lambda >0$ and $E/m \gg 1$, we can find that the velocity of 
massive particles approaches $1$ much faster than that 
in the Minkowski space-time as the energy of the particle increases. 
However, for $\lambda <0$ and $E/m \gg 1$, the difference of the velocity from $1$ 
becomes large as the mass of the particle increases. 
Note that if $|\lambda (E-m)|\ll 1$, we obtain $|\lambda m| \gg 1$ by using 
the condition $|\lambda E|\gg 1$. 
Since $E-m$ is written as $m(1/\sqrt{1-v_{M}^{2}}-1)$ in the Minkowski 
space-time, where $v_{M}$ is the velocity in the Minkowski space-time, 
we can rewrite the condition $|\lambda (E-m)|\ll 1$ as 
$|\lambda m(1/\sqrt{1-v_{M}^{2}}-1)|\ll 1$, which leads to 
$v_{M}\ll 1$ because of $|\lambda m| \gg 1$. Then, we find $v_{l}\simeq 
v_{M}\sqrt{2\lambda m}$ and $v_{l}\simeq e^{\lambda m}v_{M}\sqrt{-2\lambda m}$ 
for $\lambda >0$ and for $\lambda <0$, respectively. 
Thus, we find that $v_{l}$ for the case $|\lambda m|\gg 1$ is quite different 
from $v_{M}$, which is a good approximation for describing a velocity of 
macroscopic bodies in our world under the conditions we are considering. 
To describe a velocity of macroscopic bodies in the 
$\kappa$-Minkowski space-time, we must consider carefully what are the energy 
and the momentum, since these quantities are obtained by a total sum of 
those of elementary particles according to the addition law (\ref{add}). 
The above discrepancy may be explained in this reason. 
Below, we only consider elementary particles and restrict the discussion 
to the case $|\lambda m|\ll 1$. 

%%%%%%%%%%%%%%%%
\begin{table}[htbp]
\begin{center}
\caption{Approximation of the velocity $v_{l}$ in the case $|\lambda E|\gg 1$ 
\label{table}   }
\tabcolsep=5mm
\begin{tabular}{p{1cm}|p{4.5cm}|p{3.5cm}}
                        &$E/m\gg 1$                          & $|\lambda (E-m)|\ll 1$   \\
\hline            
$\lambda >0$  &$1+e^{-2\lambda E}\left[\frac{1}{2}-\cosh^{2}(\lambda m)\right]$  
&$2\sqrt{\lambda (E-m)}$   \\
$\lambda <0$  &$1/\cosh (\lambda m)$ &$2e^{\lambda m}\sqrt{\lambda (m-E)}$ \\
\end{tabular}
\end{center}
\end{table}
%%%%%%%%%%%%%%%%%%%

%%%%%%%%%%%%%%%%%%%%%%%%%%%%%%%%%%%%%%%
%%%%%%%%%%%%%%%%%%%%%%%%%%%%%%%%%%%%%%%
\section{Measurements of the effective scale of
``Quantum Gravity'' by massive particles}
%%%%%%%%%%%%%%%%%%%%%%%%%%%%%%%%%%%%%%%
%%%%%%%%%%%%%%%%%%%%%%%%%%%%%%%%%%%%%%%

In this section, we compare $v_{l}$ with $v_{MDR}$
and discuss the possibility of detection of effective scale of quantum
gravity by observations and experiments. The behavior of the velocities is quite
different depending on the mass and energy of the particle.
Hence, we consider two
limiting cases: (i) the ``relativistic'' case ($m \ll E$) and
(ii) ``non-relativistic'' case ($m \approx E$)\cite{chuu}.

In the relativistic case, $m \ll E$, and under the assumptions, $E \ll
E_{QG}$ and $E \ll |\lambda^{-1}|$, $v_{MDR}$ and $v_{l}$ are
%%%%%%%%%%%%%%%
\begin{eqnarray}
v_{MDR} &\approx & 1-\frac{1}{2}\left(\frac{m}{E}\right)^{2}
-\frac{\xi (n-1)}{2}\left(\frac{E}{E_{QG}}\right)^{n-2},
\label{first2LIV}\\
v_{l}&\approx &1-\frac{1}{2}\left(\frac{m}{E}\right)^{2}
+\frac{\lambda m^{2}}{2 E},
\label{first2-kappa}
\end{eqnarray}
%%%%%%%%%%%%%%%
at the lowest order of $m/E$ and $E/E_{QG}$ in the MDR models and $\lambda E$ 
in the $\kappa$-Minkowski space-time, respectively. 
When $m=0$ in the MDR models, $E_{QG}$ can be constrained by the $\gamma$-rays 
from the Mk 421 as mentioned in Sec. II. 
However, since $v_{l}=1$ for massless particles, 
(we can confirm this is also true for all order of $\lambda m$ and
$\lambda E$), $\lambda$ is not constrained by  massless particles.
This is an important result since the result notices us that there are a wide variety 
of candidates for the
theory of quantum gravity, some of which the scale of quantum gravity is not 
constrained by  present observations. The situation changes for massive particles
since the lowest order correction appears in the coupled form 
with the mass of the particle in the $\kappa$-Minkowski space-time, 
while that of the MDR models 
does not depend on the mass of the particle. 

First, we consider neutrinos from supernovae with energy
$E_{\nu} \sim 10^{10}$ eV to detect space-time noncommutativity.
We assume that the mass of an electron neutrino
and all the parameters necessary to
describe neutrino physics are determined
by other experiments and observations,
and use the delay of the arrival time between the neutrinos and gravitational
waves to evaluate the scale of quantum gravity. 
In this case, the delay of the arrival time is 
%%%%%%%%%%%%%%%
\begin{eqnarray}
\delta t\approx\frac{Lm^{2}_{\nu}}{2E_{\nu}}
\left(\frac{1}{E_{\nu}}+\lambda\right).
\label{time}
\end{eqnarray}
%%%%%%%%%%%%%%%
Since neutrinos are emitted continuously during about $10$ s, 
it is impossible to determine the time when the neutrino is emitted 
more accurate than that time scale. For this reason, 
$\delta t \gsim 10$ s is necessary to detect the effect of quantum gravity. 
As for $\lambda$, since there is no restriction from
the arrival time analysis of $\gamma$-rays, $\lambda$ may
take a large value. However, by considering reaction processes by collider 
experiments, we can restrict $|\lambda |\lsim 10^{-12}$ eV$^{-1}$ 
since the threshold of the reaction will change drastically for 
$|\lambda |>1/E_{th}$, where $E_{th}$ is the threshold energy 
in the Minkowski space-time\cite{Tamaki}.
Then, $L$ becomes far longer than the horizon scale in the present universe 
even if $|\lambda |=10^{-12}$ eV$^{-1}$.
Thus, it is difficult to detect this effect in this phenomena. 

Neutrinos from  $\gamma$-ray bursts in fireball models
have a different energy scale. In the bursts, neutrinos with energy
$\sim 10^{14}$ eV and $\gamma$-rays are expected to be radiated
away in $\sim 1$ s\cite{Bahcall}. We show that we cannot detect 
space-time noncommutativity even if we neglect the dissipation 
of the $\gamma$-ray. In the $E\gg 1/|\lambda|$ case, we can evaluate the delay of 
the arrival time of neutrinos compared with the
$\gamma$-rays from Table \ref{table} as  
%%%%%%%%%%%%%%%
\begin{eqnarray}
\delta t&\approx &\frac{L}{2}e^{-2\lambda E}\left[1+
2(\lambda m_{\nu})^{2}\right]\ \ {\rm for}\ \ \lambda >0, \\ 
\delta t&\approx &\frac{L}{2}(\lambda m_{\nu})^{2}\ \ {\rm for}\ \ \lambda <0,
\label{time2}
\end{eqnarray}
%%%%%%%%%%%%%%%
where we have used the conditions $E/m\gg 1$ and $|\lambda m|\ll 1$.
If we assume $\delta t \sim 1$ s and $|\lambda |=10^{-12}$ eV$^{-1}$,
the path of the particle's travel becomes far longer than the horizon scale 
in the present universe. 
In the $E\sim 1/|\lambda|$ case, the arrival time delay
cannot be described in a simple way. There is, however,
no qualitative difference from the above case.
Hence, it is difficult to detect space-time noncommutativity by this method.

%--------------------
Next, we examine the non-relativistic case,
$m\sim E\ll E_{QG}$ (or $|\lambda^{-1}|$). The velocity 
in each model are
%%%%%%%%%%%%%%%
\begin{eqnarray}
v_{MDR}&\approx &\sqrt{1-\left(\frac{m}{E}\right)^{2}}\times
\nonumber \\
&& \;\;\;\;\;
\left[1+\frac{\xi}{2}
\frac{E^{2}(1-n)+nm^{2}}{E^{2}-m^{2}}
\left(\frac{E}{E_{QG}}\right)^{n-2}\right],
\label{firstLIV}\\
v_{l}&\approx &\sqrt{1-\left(\frac{m}{E}\right)^{2}}
\left(1+\frac{\lambda m^{2}}{2E}\right).
\label{first-kappa}
\end{eqnarray}
%%%%%%%%%%%%%%%
Note that the absolute value of the correction for the velocity in the 
$\kappa$-Minkowski space-time 
decreases with the energy, while that in the MDR model increases. 
Although, in the low-energy limit, 
the dispersion relation in the $\kappa$-Minkowski space-time has the same 
form as that in the MDR models, the correction for the velocity is quite different. 

Because of the above difference in the correction terms, there is
a possibility that the evidence for space-time
noncommutativity can be detected in use of the low-energy
massive particles. Here, we consider the ultra-cold neutrons
with energy $E_n-m_n \sim 10^{-2}$ eV\cite{Dubbers}.
Since the mass of a neutron $m_n$ is measured with high
accuracy, we can estimate the time interval in which the neutron
travels the interval $L$ in the Minkowski space-time. 
If a time lag is obtained in an experiment,
it can be interpreted as the effect of space-time
noncommutativity. This time lag is calculated in the 
$\kappa$-Minkowski space-time as
%%%%%%%%%%%%%%%
\begin{eqnarray}
\delta t=\frac{L}{v_{l}}-\frac{L}{v_{M}}\approx
\frac{L}{v_{M}}\frac{\lambda m_n^{2}}{2E_n}.
\label{difference}
\end{eqnarray}
%%%%%%%%%%%%%%%
By substituting the value of
the apparatus\cite{Rich}, $L\sim 100$ m, we have
%%%%%%%%%%%%%%%
\begin{eqnarray}
\delta t\sim 10^{-1}\lambda m_n.
\label{value}
\end{eqnarray}
%%%%%%%%%%%%%%%
If the resolution for the measurement of the time lag is
$\sim 10^{-10}$ s and $|\lambda |\agt 10^{-18}$ eV$^{-1}$, we can detect
space-time noncommutativity.

%%%%%%%%%%%%%%%%%%%%%%%%%%%%%%%%%%%%%%%
%%%%%%%%%%%%%%%%%%%%%%%%%%%%%%%%%%%%%%%
\section{Conclusion}
%%%%%%%%%%%%%%%%%%%%%%%%%%%%%%%%%%%%%%%
%%%%%%%%%%%%%%%%%%%%%%%%%%%%%%%%%%%%%%%

We have investigated what are the qualitative differences of the velocity 
formula in the $\kappa$-Minkowski space-time from that in the MDR models. 
Most of the previous papers had adopted the MDR models since the MDR models are 
among the simplest ones of quantum gravity. 
However, many of the MDR models do not have physical foundation in how 
the correction terms of naturally arise in the dispersion relation. 
For example, since the usual Lorentz transformation 
had been used in the previous work, one could not have avoided the existence of a 
preferred frame as a result. Since we have taken the standpoint that 
the existence of a preferred frame is not favorable, we have considered 
the $\kappa$-Minkowski space-time where the deformed Lorentz transformation 
and the deformed dispersion relation arise as a result of the deformation 
quantization. 

We have found that since massless particles move in a constant speed
in the $\kappa$-Minkowski space-time, the arrival time analyses by $\gamma$-rays
are not capable to detect the difference from the Minkowski space-time. 
This example shows that it is difficult to constrain all kinds of 
Lorentz invariance by a single experiment. 
Therefore, we need to investigate specific models individually.
We have also considered the possibility to detect space-time noncommutativity
by low-energy massive particles. In our model, if
the resolution for the measurement of the time lag is given by 
$\sim 10^{-10}$ s, it {\it is} possible to constrain $\lambda$ to 
$|\lambda |\agt 10^{-18}$ eV$^{-1}$.
Although these features had not been investigated so far, it may be important.

%%%%%%%%%%%%%%%%%%%%%%%%%%%%%%%%%%%%%%%
%%%%%%%%%%%%%%%%%%%%%%%%%%%%%%%%%%%%%%%
\section*{Acknowledgments:}
%%%%%%%%%%%%%%%%%%%%%%%%%%%%%%%%%%%%%%%
%%%%%%%%%%%%%%%%%%%%%%%%%%%%%%%%%%%%%%%
Special thanks to Kei-ichi Maeda for continuous encouragement. 
This work was supported partly by the Grant-in-Aid (No.05540) from
the Japanese Ministry of Education, Culture, Sports, Science and Technology,
and partly by the Waseda University Grant for Special Research Projects.

%%%%%%%%%%%%%%%%%%%%%%%%%%%%%%%%%%%%%%%


\begin{thebibliography}{99}
\bibitem{footnote2}
This is not a Gaussian wave packet. However, it is 
sufficient to obtain a group velocity. The extension for more 
general wave packet will be straight forward.
\bibitem{chuu}
In the MDR models and in the $\kappa$-Minkowski space-time,
it is possible that the particle moves very slowly 
(fast) even if the condition $m \ll E$ ($m \approx E$)
is satisfied.
\end{thebibliography}
\end{document}